\documentclass{aa}
\usepackage{txfonts}
\usepackage{multirow}
\usepackage{graphicx}
\usepackage{natbib}
\begin{document}
\title{A study of the neglected Galactic \ion{H}{ii} region NGC\,2579 and its 
companion ESO\,370-9}
\subtitle{}
\author{ 
      M.~V.~F.~Copetti\inst{1} 
  \and 
      V.~A.~Oliveira\inst{1}
  \and 
      R.~Riffel\inst{2}
  \and 
      H.~O.~Casta\~neda\inst{3}
  \and D. Sanmartim\inst{1}}
\offprints{M.~V.~F.~ Copetti \\ \email{mvfc@lana.ccne.ufsm.br}}
\institute{
    Laborat\'orio de An\'alise Num\'erica e Astrof\'{\i}sica,
    Departamento de Matem\'atica, e
    Programa de P\'os-Gradua\c{c}\~ao em F\'{\i}sica,
    Universidade Federal de Santa Maria,
    97119-900 Santa Maria, RS, Brazil. 
  \and
    Departamento de Astronomia,     
    Universidade Federal do Rio Grande do Sul.    
    Av. Bento Gon\c{c}alves 9500, Porto Alegre, RS, Brazil
  \and
    Instituto de Astrofísica de Canarias, La Laguna, Tenerife, 
    E-38200 Canary Islands, Spain   
          }
\date{Received 15 May 2007 / Accepted 15 June 2007}

%
%
\abstract
%
%
{ 
The Galactic \ion{H}{ii} region NGC\,2579 has stayed undeservedly unexplored due 
to identification problems which persisted until recently. Both NGC\,2579 and 
its companion ESO\,370-9 have been misclassified as planetary or reflection 
nebula, confused with each other and with other objects. Due to its high surface brightness, high excitation, angular size of few arcminutes and 
relatively low interstellar extinction, NGC\,2579 is an ideal object for 
investigations in the optical range. Located in the outer Galaxy, NGC\,2579 is 
an excellent object for studying the Galactic chemical abundance gradients.
}
%
%
{
To present the first comprehensive observational study on the nebular and 
stellar properties of NGC\,2579 and ESO\,370-9, including the determination of 
electron temperature, density structure, chemical composition, kinematics, 
distance, and the identification and spectral classification of the ionizing 
stars, and to discuss the nature of ESO\,370-9.
}
%
%
{
Long slit spectrophotometric data in the optical range were used to derive the 
nebular electron temperature, density and chemical abundances and for the 
spectral classification of the ionizing star candidates. H$\alpha$ and $UBV$ CCD 
photometry was carried out to derive stellar distances from spectroscopic 
parallax and to measure the ionizing photon flux. 
}
%
%
{
The chemical abundances of He, N, O, Ne, S, Cl, and Ar were obtained. Maps of 
electron density and radial velocity with a spatial resolution of 5\arcsec 
$\times$ 5\arcsec\ were composed from long slit spectra taken at different 
declinations. Three O stars classified as O5\,V, O6.5\,V, and O8\,V were found 
responsible for the ionization of NGC\,2579, while ESO\,370-9 is ionized by a 
single O8.5\,V star. The estimated mass of ionized gas of $\approx$ 25 $M_\odot$ 
indicates that ESO\,370-9 is not a planetary nebula, but a small \ion{H}{ii} 
region. A photometric distance of 7.6 $\pm$ 0.9 kpc and a kinematic 
distance of 7.4 $\pm$ 1.4 kpc were obtained for both objects. At the 
galactocentric distance of 12.8 $\pm$ 0.7 kpc, NGC\,2579 is one of the most 
distant Galactic \ion{H}{ii} regions for which direct abundance determinations 
have been accomplished.
}
%
%
{}
\keywords{ISM: \ion{H}{ii} regions -- ISM:planetary nebulae}
\maketitle
%
%
\section{Introduction}
Due to observational difficulties, the knowledge acquisition in many research 
areas of astrophysics has still relied largely on data obtained from a small 
group of characteristic objects. This is certainly the case of the optical 
studies of Galactic \ion{H}{ii} regions. In this field, the Orion Nebula is by 
far the most studied object. In fact, the number of papers on this object is 
comparable to the total number of optical studies of all other Galactic 
\ion{H}{ii} regions. Even the standard procedure of measuring the electron 
temperature in Galactic \ion{H}{ii} regions from optical emission line ratios, 
such as [\ion{O}{iii}]\,($\lambda 4959 + \lambda 5007)/\lambda 4363$ and 
[\ion{N}{ii}]\,($\lambda 6548 + \lambda 6584)/\lambda 5755$, until now could 
only be accomplished in a small sample of a dozen or so objects. So, the 
addition of new members to the selected list of well studied (or easy to 
observe) objects is welcome, especially because the \ion{H}{ii} regions are good 
tracers of the chemical abundance variation across the disk of the Galaxy.

The Galactic \ion{H}{ii} region \object{NGC\,2579} (centred at $\alpha=$ 
08:20:54.8, $\delta=-$36:12:59.9, J2000), because of its high surface 
brightness, angular size of few arcminutes and relatively low interstellar 
extinction, is an ideal object for investigations in the optical range, but 
stays undeservedly unexplored due to identification problems which persisted 
until recently. Despite its description as a ``double star in a pretty small 
nebula among 70 stars'' and sufficiently precise coordinates in the New 
General catalogue \citep{Dreyer 1888}, NGC\,2579 has been confused with other 
nearby (and not so nearby) objects, as the open clusters \object{NGC\,2580} and 
\object{AH03 J0822-36} \citep{Archinal & Hynes 2003}, which show no clear signs 
of nebulosity. The only $UBV$ photometric study found in the literature of the 
stars claimed to be in NGC\,2579, by \citet{Lindoff 1968}, is in fact on stars 
pertaining to the cluster AH03 J0822-36, which lies approximately 20\arcmin\ to 
the Southeast of the supposed object. NGC\,2579 is sometimes incorrectly 
associated with the nebula \object{RCW\,20}. Correct identifications of 
NGC\,2579 are with \object{Gum\,11} \citep{Gum 1955} and \object{BBW\,138} 
\citep{Brand et al. 1986}. 

NGC\,2579 has been misclassified as a reflection nebula (e.g., in the SIMBAD 
database) because of the association with the objects \object{VdBH\,13a}, 
\object{VdBH\,13b}, and \object{VdBH\,13c} wrongly considered of this kind. 
These objects are in fact stars towards NGC\,2579 listed in the catalogue of 
``southern stars embedded in nebulosity" by \citet{van den Bergh & Herbst 1975}, 
which was a result of a survey of southern reflection nebulae conducted by these 
authors. However, not all of the stars in the catalogue are really associated 
with reflection nebulae. As another example of the confusion involving the 
identification of NGC\,2579, these authors mistakenly associated VdBH\,13a, b, 
c with \object{NGC\,2580}, an open cluster more than 6 degrees away from the 
observed position.

NGC\,2579 was ``rediscovered" as the emission nebula \object{Ns\,238} in the 
objective prism survey by \citet{Nordstrom 1975} and classified as a probable 
planetary nebula. It is identified as PN G254.6+00.2 in the Strasbourg-ESO 
catalogue of Galactic planetary nebulae \citep{Acker et al. 1992}. However, 
doubts about the planetary nebula nature of this object were raised 
\citep{Kimeswenger 2001}, based on the morphology, IRAS colors and total far 
infrared flux.

To complicate the matter, NGC\,2579 has a very close companion, the object 
\object{ESO\,370-9}, a small ($\sim$ 1\arcmin) roughly elliptical ringed nebula 
with a star in the middle ($\alpha=$ 08:20:56.75, $\delta=-$36:13:46.9, J2000). 
It was discovered and classified as a planetary nebula candidate by \citet[the 
original identification was 370-PN?09]{Lauberts et al. 1981}. In the SIMBAD 
database ESO\,370-9 is classified as a reflection nebula, because of its 
association with VdBH\,13c, which is its central star. It is also confused with 
the misclassified planetary nebula PN G254.6+00.2 (in fact with \ion{H}{ii} 
region NGC\,2579).

\begin{figure}
\centering
\resizebox{\hsize}{!}{\includegraphics*[angle=-90]{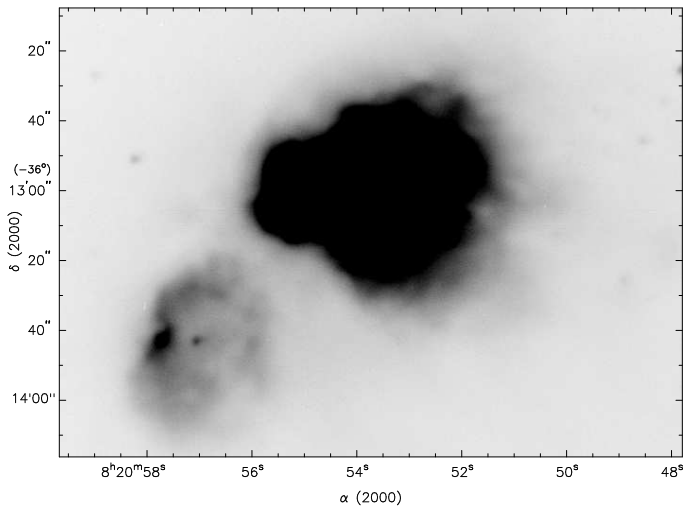}}
\caption{H$\alpha$ image of NGC\,2579 (brighter nebula) and ESO\,370-9} 
\label{image}
\end{figure}

In this paper, we present the first comprehensive optical study on the 
nebular and stellar properties of the Galactic \ion{H}{ii} region NGC\,2579 and 
its companion ESO\,370-9, including the determination of electron temperature, 
density structure, chemical composition, kinematics, distance, and the 
identification and spectral classification of its ionizing stars. We also 
discuss the nature of ESO\,370-9. In Fig.~\ref{image} we present an H$\alpha$ 
image of the area of NGC\,2579 and ESO\,370-9, which shows that these two nebula 
are distinct objects.

%
%

%
%
\section{Observations and data reductions}
\label{observations}

\subsection{Spectroscopy}
Long slit spectrophotometric observations were carried out with the Boller \& 
Chivens spectrograph attached to the 1.52 m telescope at the European Southern 
Observatory (ESO), La Silla, Chile, and with the Cassegrain spectrograph 
attached to the 1.6 m telescope at Observat\'orio do Pico dos Dias (OPD), 
Bras\'opolis, Brazil. At ESO we used a Loral CCD of 2688 $\times$ 512 pixels and 
a grid of 1200 grooves mm$^{-1}$, resulting in a spatial scale of $1.64\arcsec$  
pxl$^{-1}$ (after the rebinning of each two contiguous CCD rows to increase the 
signal-to-noise ratio), a spectral dispersion of 1.0 \AA\, pxl$^{-1}$, and a 
resolution of 2.9 \AA. At the OPD we used a SITe CCD of 2048 $\times$ 2048 
pixels, resulting in a spatial scale of $0.56 \arcsec$ pxl$^{-1}$, and two 
different grids. Spectra in the range of 4000 to 7750 \AA, with dispersion of 
2.4 \AA\, pxl$^{-1}$ and resolution of about 8 \AA, were obtained with a grid of 
300 mm$^{-1}$, while a grid of 1200 grooves mm$^{-1}$ was used to obtain 
spectra in the 4030--4990 \AA\ and 6000--7000 \AA\ ranges, with dispersion of 
0.5 \AA\, pxl$^{-1}$ and mean resolution of 2.7 \AA. Table 
\ref{journal_spec_obs} presents the journal of observations. The columns are the 
declination offset $\Delta\delta$ of the slit relative to reference star DENIS 
J082054.8-361258 ($\alpha=$ 08:20:54.86, $\delta=-$36:12:58.9, epoch J2000), the 
observatory and telescope used, the spectral range, and the number and time of 
the exposures.

The observation routine followed usual procedures. Dome flat-field exposures 
were taken at the beginning and at the end of each night. About 30 bias frames 
were made per night. Spectrophotometric standard stars were observed for flux 
calibration. Spectra of a He--Ar--Ne lamp were taken before and after each 
object exposure for wavelength calibration. Exposure times were limited to 20 
minutes to reduce the effects of cosmic rays. 

%
%
\begin{table}
\caption{Journal of spectroscopic observations}
\label{journal_spec_obs}
\begin{tabular}{lllll}
\hline
\noalign{\smallskip}
$\Delta\delta$ & Obs./Tel. & Date & Wavelength  & Exposure \\
       &                   &      & range (\AA) & (s)      \\
\noalign{\smallskip}
\hline
\noalign{\smallskip}
0\arcsec     & ESO 1.52 m & 2002 Dec 30 & 3300--5100 & $3 \times 1200$ \\
             & ESO 1.52 m & 2002 Dec 31 & 4630--6750 & $3 \times 1200$ \\

8\arcsec\,N  & ESO 1.52 m & 2002 Dec 30 & 3300--5100 & $1 \times 1200$ \\
             & ESO 1.52 m & 2003 Jan 01 & 3300--5100 & $2 \times 1200$ \\
             & ESO 1.52 m & 2002 Dec 31 & 4630--6750 & $2 \times 600 $  \\

7\arcsec\,S  & ESO 1.52 m & 2003 Jan 01 & 3300--5100 & $3 \times 1200$ \\
             & ESO 1.52 m & 2002 Dec 31 & 4630--6750 & $3 \times 600 $ \\

44\arcsec\,S & OPD 1.60 m & 2006 Apr 20 & 4030--4990 & $3 \times 1200$ \\
             & OPD 1.60 m & 2006 Apr 24 & 4000--7750 & $3 \times 1200$ \\
\noalign{\smallskip}
 0\arcsec     & OPD 1.60 m & 2004 Apr 22 & 6000--7000 & $2 \times 1200$ \\
 5\arcsec\,N  & OPD 1.60 m & 2004 Nov 08 & 6000--7000 & $1 \times 600$  \\
10\arcsec\,N  & OPD 1.60 m & 2004 Apr 22 & 6000--7000 & $1 \times 1200$ \\
15\arcsec\,N  & OPD 1.60 m & 2004 Apr 22 & 6000--7000 & $2 \times 1200$ \\
20\arcsec\,N  & OPD 1.60 m & 2004 Nov 08 & 6000--7000 & $2 \times 600$  \\
25\arcsec\,N  & OPD 1.60 m & 2004 Apr 22 & 6000--7000 & $1 \times 1200$ \\
              & OPD 1.60 m & 2004 Nov 08 & 6000--7000 & $1 \times 1200$ \\
30\arcsec\,N  & OPD 1.60 m & 2004 Apr 22 & 6000--7000 & $1 \times 1200$ \\
              & OPD 1.60 m & 2004 Nov 08 & 6000--7000 & $1 \times 1200$ \\
 5\arcsec\,S  & OPD 1.60 m & 2004 Apr 23 & 6000--7000 & $1 \times 1200$ \\
10\arcsec\,S  & OPD 1.60 m & 2004 Apr 23 & 6000--7000 & $1 \times 1200$ \\
15\arcsec\,S  & OPD 1.60 m & 2004 Apr 23 & 6000--7000 & $1 \times 1200$ \\
20\arcsec\,S  & OPD 1.60 m & 2004 Apr 23 & 6000--7000 & $1 \times 1200$ \\
25\arcsec\,S  & OPD 1.60 m & 2004 Apr 23 & 6000--7000 & $1 \times 1200$ \\
30\arcsec\,S  & OPD 1.60 m & 2004 Apr 23 & 6000--7000 & $1 \times 1200$ \\
35\arcsec\,S  & OPD 1.60 m & 2004 Apr 23 & 6000--7000 & $1 \times 1200$ \\
40\arcsec\,S  & OPD 1.60 m & 2004 Apr 23 & 6000--7000 & $1 \times 1200$ \\
\noalign{\smallskip}
\hline
\end{tabular}
\end{table}

The slits used had entrances on the plane of sky of 2$\arcsec \times 250 
\arcsec$ for the observations at ESO and 1.5$\arcsec \times 320 \arcsec$ for the 
observations at OPD, and they were aligned along the east-west direction. For 
the identification and spectral classification of the ionizing star (see 
Sect.~\ref{ionizing stars}), in some of the observations, the slit was set 
passing through some plausible ionizing star candidate. Spectra of the four 
brightest stars on the image of NGC\,2579 and of the central star in ESO\,370-9 
were obtained with slit declination corresponding to $\Delta\delta$ = 0\arcsec, 
8\arcsec\,N, 7\arcsec\,S, and 44\arcsec\,S. These spectra cover most of the 
optical range and were also used for nebular abundance determinations (see 
Sect.~\ref{abundance analysis}). In the other observations, the slit was 
positioned  at 15 different and equally spaced declinations separated by 
5\arcsec. The 2D spectra obtained in the range 6000--7000 \AA, sampling the 
whole nebula NGC\,2579 and part of ESO\,370-9, were used to produce maps of 
electron density and radial velocity (see Sect.~\ref{electron density and 
velocity maps}). 

The data reduction (bias correction, flat-fielding, cosmic rays cleaning, 
wavelength and flux calibration, 1D spectrum extraction) followed standard 
procedures and was made with the IRAF software. From the combined spectra 
acquired at $\Delta\delta$ = 0\arcsec, 8\arcsec\,N, 7\arcsec\,S, and 
44\arcsec\,S., five 1D stellar spectra and four nebular spectra integrated along 
the slit were obtained. Due to the great and irregular surface brightness of the 
nebula, the background subtraction for the stellar spectra was difficult and 
many attempts had to be made to eliminate the contamination of the nebular 
emission lines. From each of the 2D spectra obtained at equally spaced 
declinations, we extracted a series of 1D spectra from contiguous sectors of 
5\arcsec\ of length along the slit axis (east-west direction). To secure 1D 
spectra from sectors along the same north-south strips, we first had to define a 
fiducial position on the slit axis, by measuring in the 2D spectrum spatial 
profiles the position of the detected star and comparing them with coordinates 
obtained from direct images of the region. In this way, we extracted 750 1D 
spectra from individual sectors of size $5\arcsec \times 1.5\arcsec$ centred on 
a grid of $50 \times 15$ equally spaced points, sampling the nebula with a 
spatial resolution of $5\arcsec \times 5\arcsec$.

The emission line fluxes were obtained by Gaussian fitting of the line profile 
and by direct integration of the flux over a linear local continuum defined by 
eye carried out with the {\it splot} routine of the IRAF package. We estimated 
the error associated with the line fluxes by $\sigma^{2} = \sigma_{\rm cont}^{2} 
+ \sigma_{\rm line}^{2} + \sigma_{\rm cal}^{2} + \sigma_{\rm meth}^{2}$, where 
$\sigma_{\rm cont}$ is the error due to the continuum baseline determination, 
$\sigma_{\rm line}$ is the Poisson error of the emission line, $\sigma_{\rm 
cal}$ is the error due to the flux calibration (measured as the standard 
deviation of the residual of the fitting of the standard star calibration 
curves), and $\sigma_{\rm meth}$ is the error due to the method of integration 
of the line flux. All the line intensities were normalized to H$\beta$ and 
corrected for the effect of the interstellar extinction by comparing the 
observed ratios $\rm H\alpha/H \beta$, $\rm H\gamma/H \beta$ and $\rm H\delta/H 
\beta$ with the theoretical ones calculated by \citet{Storey & Hummer 1995} for 
an electron temperature of 10\,000 $\mathrm{K}$ and a density of 100 
$\mathrm{cm^{-3}}$. The Galactic reddening function of \citet{Kaler 1976} was 
used. In some positions, significantly different values for the logarithmic 
extinction coefficient $c(\rm H \beta)$ were obtained from the blue and red 
spectra. To mitigate possible systematic errors in the flux calibration which 
could be responsible for these differences, the intensities of the lines with 
wavelength larger than that of H$\beta$ were corrected with $c(\rm H \beta)$ 
calculated from the H$\alpha/$H$\beta$ ratio and for the other lines we used the 
mean value from the H$\gamma/$H$\beta$ and H$\delta/$H$\beta$ ratios. Table 
\ref{line intensities} lists the observed and reddening-corrected emission line 
intensities relative to H$\beta$ and the logarithmic extinction coefficient 
$c(\rm H \beta)$ from the integrated spectra obtained for NGC\,2579 at offset 
declinations $\Delta\delta$ = 0\arcsec, 8\arcsec\,N, 7\arcsec\,S (hereafter 
labelled A, B, and C, respectively), from the sum of these three spectra 
(labelled ABC), and for ESO\,370-9 from the spectrum obtained at $\Delta\delta$ 
= 44\arcsec\,S. 
%
%
\begin{table}
\caption{Journal of photometric observations}
\label{journal_phot_obs}
\begin{tabular}{llll}
\hline
\noalign{\smallskip}
Filter & Obs./Tel. & Date & Exposure (s)  \\
\noalign{\smallskip}
\hline
\noalign{\smallskip}
$U$ & SPM 1.5 m & 2000 Mar 10 & $3 \times 300$                              \\
    & SPM 1.5 m & 2000 Mar 14 & $1 \times 300$                              \\
$B$ & SPM 1.5 m & 2000 Mar 10 & $2 \times 300 + 1 \times 100$               \\
    & SPM 1.5 m & 2000 Mar 14 & $1 \times 300 + 1 \times 200$               \\
    & OPD 0.6 m & 2005 Apr 10 & $5 \times 240 + 5 \times 120$               \\
$V$ & SPM 1.5 m & 2000 Mar 10 & $1 \times 60 + 1 \times 30 + 1 \times 20$   \\
    & SPM 1.5 m & 2000 Mar 14 & $1 \times 30$                               \\
    & OPD 0.6 m & 2005 Apr 10 & $5 \times 45 + 7 \times 35 + 5 \times 25$   \\
 H$\alpha$ & SPM 1.5 m & 2000 Mar 15 & $5 \times 300$ \\ 
 Cont.     & SPM 1.5 m & 2000 Mar 15 & $4 \times 60$  \\
\noalign{\smallskip}
\hline
\end{tabular}
\end{table}

\subsection{$UBV$ photometry}

Direct images in the $UBV$ filters were obtained with the 1.5 m telescope at 
the San Pedro M\'artir Observatory (SPM), B.C., Mexico and with the 0.6 m Boller 
\& Chivens telescope at the Observat\'orio do Pico dos Dias (OPD), Bras\'opolis, 
Brazil, in March 2000 and April 2005, respectively. To avoid saturation of the 
brightest stars, multiple exposures of different times were taken in each 
filter. Several dome flat-field and bias exposures were taken at the beginning 
and at the end of each night. The $UBV$ \citet{Landolt 1992} standard star 
fields SA 99-447/438, SA 106-700 were observed at the OPD and PG 0918+029, PG 
1323-086 and Feige\,34 at SPM. About 40 secondary standard stars measured by 
\citet{Galadi-Enriquez et al. 2000} were found in these fields. The journal of 
the photometric observations is presented in Table~\ref{journal_phot_obs}. The 
seeing ranged from 1.5\arcsec\ to 2.5\arcsec\ in different nights. 

The reduction followed the standard procedure for stellar CCD photometry in a 
relatively crowded field and was performed with the IRAF/Daophot package. 
%
%
\begin{table}
\caption{Interference filters}
\label{filters}
\begin{tabular}{lllll}
\hline
\noalign{\smallskip}
Filter & $\lambda_{\rm c}$ (\AA) & Peak trans. & FWHM (\AA) & Effec. width (\AA) \\
\noalign{\smallskip}
\hline
\noalign{\smallskip}
H$\alpha$ & 6563 & 66.4\% & 9.8 & 5.4 \\
Cont.     & 6450 & 95.0\% & 127 & 114 \\
\noalign{\smallskip}
\hline
\end{tabular}
\end{table}

\subsection{H$\alpha$ photometry}
\label{Halpha photometry}

To measure the H$\alpha$ flux of the nebulae we also obtained CCD images in 
narrow-band interference filters in H$\alpha$ and near continuum with the 1.5 m 
telescope at the San Pedro M\'artir Observatory. The log of observations is in 
Table~\ref{journal_phot_obs}. The spectrophotometric standard star Feige 34 was 
observed for flux calibration. Details of the filters, including the central 
wavelength $\lambda_{\rm c}$, effective bandwidth, peak transmission, and full 
width at half maximum (FWHM), are presented in Table~\ref{filters}. After the 
standard CCD data reduction, the brightest stars were eliminated from the images 
by interpolation of nearby data points and the total counts in each filter 
within a polygonal area encompassing the nebulae were obtained. After this we 
followed a procedure similar to that described by \citet{Copetti & Dottori 1989} 
to calculate the H$\alpha$ flux from the counts, taking into account the small 
but not insignificant contribution of the nebular emission lines 
[\ion{S}{iii}]$\lambda6312$, [\ion{N}{ii}]$\lambda 6583$ and even H$\alpha$ to 
the counts in the continuum filter. We obtained H$\alpha$ fluxes (in erg 
cm$^{-2}$ s$^{-1}$) of $\log F(\mathrm{H}\alpha) = -9.68$ for NGC\,2579 and 
$\log F(\mathrm{H}\alpha) = -10.58$ for ESO\,370-9.

%
%
%

%
%
\begin{table*}[p]
\caption{Observed and reddening-corrected emission line intensities, 
$F(\lambda)$ and $I(\lambda)$, respectively (normalized to H$\beta=100$). 
Positions A, B and C correspond to $\Delta\delta=0\arcsec$, 8\arcsec\,N, and 7\arcsec\,S, respectively. 
The spectrum of ESO\,370-9 was obtained at $\Delta\delta=44\arcsec$\,S}
\label{line intensities}
\begin{tabular}{l@{~~}r@{~}l@{~~~}
       r@{~~}r@{~~}r@{}c@{~~~~}
       r@{~~}r@{~~}r@{}c@{~~~~}
       r@{~~}r@{~~}r@{}c@{~~~~}
       r@{~~}r@{~~}r@{}c@{~~~~}
       r@{~~}r@{~~}r@{}r@{}}
\hline
\noalign{\smallskip}
& & & \multicolumn{3}{c}{NGC\,2579, A} & 
    & \multicolumn{3}{c}{NGC\,2579, B} & 
    & \multicolumn{3}{c}{NGC\,2579, C} & 
    & \multicolumn{3}{c}{NGC\,2579, ABC} & 
    & \multicolumn{3}{c}{ESO\,370-9} \\ 
\cline{4-6} \cline{8-10} \cline{12-14} \cline{16-18} \cline{20-22}
\noalign{\smallskip}
$\lambda_0\ (\AA)$ & Ion & 
                    & $F(\lambda)~~$ & $I(\lambda)~~$ & error & 
                    & $F(\lambda)~~$ & $I(\lambda)~~$ & error &
                    & $F(\lambda)~~$ & $I(\lambda)~~$ & error &
                    & $F(\lambda)~~$ & $I(\lambda)~~$ & error &
                    & $F(\lambda)~~$ & $I(\lambda)~~$ & error \\
\noalign{\smallskip}
\hline
\noalign{\smallskip}

3726.03 & [\ion{O}{ii}]     & \multirow{2}{*}{\big\}} 
                            & \multirow{2}{*}{ 60.20} & \multirow{2}{*}{ 159.30} &  \multirow{2}{*}{  6} &
                            & \multirow{2}{*}{ 79.44} & \multirow{2}{*}{ 171.20} &  \multirow{2}{*}{  6} &
                            & \multirow{2}{*}{ 76.76} & \multirow{2}{*}{ 143.70} &  \multirow{2}{*}{  6} &
                            & \multirow{2}{*}{ 68.29} & \multirow{2}{*}{ 156.00} &  \multirow{2}{*}{  5} &
                            & \multirow{2}{*}{   ---} & \multirow{2}{*}{    ---} &  \multirow{2}{*}{---} &
                            \\

3728.82 & [\ion{O}{ii}]     \\   

3750.15 &  \ion{H}{i}       &    
                            &    1.48 &   3.83 &  13 &
                            &    1.43 &   3.03 &  13 &
                            &    2.23 &   4.12 &  14 & 
                            &    1.65 &   3.70 &   9 &
                            &     --- &    --- & --- &
                            \\  

3770.63 &  \ion{H}{i}       &    
                            &    1.69 &   4.29 &  11 &  
                            &    2.22 &   4.63 &  10 &
                            &    2.07 &   3.77 &  11 & 
                            &    1.81 &   3.99 &  12 &
                            &     --- &    --- & --- &
                            \\  

3797.90 &  \ion{H}{i}       &      
                            &    2.19 &   5.43 &   9 &
                            &    2.73 &   5.58 &   8 &
                            &    3.57 &   6.40 &   9 & 
                            &    2.61 &   5.64 &   7 &
                            &     --- &    --- & --- &
                            \\  

3835.39 &  \ion{H}{i}       &  
                            &    2.94 &   7.04 &   7 &
                            &    3.67 &   7.30 &   5 &
                            &    4.16 &   7.30 &   5 & 
                            &    3.46 &   7.25 &   4 &
                            &     --- &    --- & --- &
                            \\  

3869.06 & [\ion{Ne}{iii}]   &  
                            &   10.12 &  23.48 &   6 &
                            &   10.95 &  21.26 &   6 &
                            &   13.29 &  22.85 &   6 &
                            &   11.09 &  22.65 &   6 &
                            &     --- &    --- & --- &
                            \\    

3888.65 &  \ion{He}{i}      & \multirow{2}{*}{\big\}} 
                            & \multirow{2}{*}{  8.54} & \multirow{2}{*}{  19.46} &  \multirow{2}{*}{  5} &
                            & \multirow{2}{*}{ 10.37} & \multirow{2}{*}{  19.85} &  \multirow{2}{*}{  5} &
                            & \multirow{2}{*}{ 11.55} & \multirow{2}{*}{  19.63} &  \multirow{2}{*}{  5} &
                            & \multirow{2}{*}{  9.66} & \multirow{2}{*}{  19.43} &  \multirow{2}{*}{  5} &
                            & \multirow{2}{*}{   ---} & \multirow{2}{*}{    ---} &  \multirow{2}{*}{---} &
                            \\
3889.05 &  \ion{H}{i}       \\   

3967.79 & [\ion{Ne}{iii}]   & \multirow{2}{*}{\big\}} 
                            & \multirow{2}{*}{ 10.86} & \multirow{2}{*}{  23.04} &  \multirow{2}{*}{  4} &
                            & \multirow{2}{*}{ 13.42} & \multirow{2}{*}{  24.28} &  \multirow{2}{*}{  3} &
                            & \multirow{2}{*}{ 15.50} & \multirow{2}{*}{  25.16} &  \multirow{2}{*}{  3} &
                            & \multirow{2}{*}{ 12.75} & \multirow{2}{*}{  24.13} &  \multirow{2}{*}{  3} &
                            & \multirow{2}{*}{   ---} & \multirow{2}{*}{    ---} &  \multirow{2}{*}{---} &
                            \\

3970.07 &  \ion{H}{i}       \\  
 
4026.19 &  \ion{He}{i}      &    
                            &    0.80 &   1.61 &  13 &
                            &    1.09 &   1.89 &  13 &
                            &    1.49 &   2.34 &   9 & 
                            &    1.14 &   2.06 &   8 &
                            &     --- &    --- & --- &
                            \\  

4068.60 & [\ion{S}{ii}]     &    
                            &    0.31 &   0.60 &  25 &   
                            &    0.34 &   0.57 &  46 &
                            &    0.74 &   1.13 &  14 & 
                            &    0.41 &   0.72 &  22 &
                            &     --- &    --- & --- &
                            \\  

4101.74 &  \ion{H}{i}       & 
                            &   13.43 &  25.31 &   5 &
                            &   15.74 &  25.94 &   4 &
                            &   17.35 &  26.09 &   4 & 
                            &   15.06 &  25.78 &   3 &
                            &   12.78 &  25.28 &   9 &
                            \\

4267.26 &  \ion{C}{ii}      &    
                            &    0.20 &   0.33 &  24 &  
                            &    0.35 &   0.52 &  21 &
                            &    0.37 &   0.51 &  24 & 
                            &    0.28 &   0.43 &  17 &
                            &     --- &    --- & --- &
                            \\  

4319.63 &  \ion{O}{ii}      \\        

4340.47 &  \ion{H}{i}       &   
                            &   30.83 &  47.57 &   6 &
                            &   33.19 &  46.72 &   6 &
                            &   35.21 &  46.56 &   6 & 
                            &   32.50 &  46.95 &   6 &
                            &   29.78 &  47.50 &   5 &
                            \\ 
                            
4363.21 & [\ion{O}{iii}]    &    
                            &    1.41 &   2.14 &   7 &
                            &    1.30 &   1.80 &   6 &
                            &    1.12 &   1.46 &   7 & 
                            &    1.20 &   1.71 &   5 &
                            &     --- &    --- & --- &
                            \\  

4387.93 &  \ion{He}{i}      &    
                            &    0.36 &   0.53 &  12 &
                            &    0.42 &   0.57 &  18 &
                            &    0.64 &   0.83 &   8 & 
                            &    0.44 &   0.62 &  12 &
                            &     --- &    --- & --- &
                            \\  

4471.48 &  \ion{He}{i}      &   
                            &    3.19 &   4.42 &   6 &
                            &    3.42 &   4.43 &   4 &
                            &    3.29 &   4.06 &   4 & 
                            &    3.22 &   4.30 &   3 &
                            &    2.36 &   3.35 &  14 &
                            \\
   
4658.05 & [\ion{Fe}{iii}]   & \multirow{2}{*}{\big\}} 
                            & \multirow{2}{*}{  0.35} & \multirow{2}{*}{   0.41} &  \multirow{2}{*}{ 30} &
                            & \multirow{2}{*}{  0.51} & \multirow{2}{*}{   0.58} &  \multirow{2}{*}{ 20} &
                            & \multirow{2}{*}{  0.30} & \multirow{2}{*}{   0.33} &  \multirow{2}{*}{ 19} &
                            & \multirow{2}{*}{  0.42} & \multirow{2}{*}{   0.48} &  \multirow{2}{*}{ 18} &
                            & \multirow{2}{*}{   ---} & \multirow{2}{*}{    ---} &  \multirow{2}{*}{---} &
                            \\

4661.63 &  \ion{O}{ii}      \\  

4713.14 &  \ion{He}{i}      &    
                            &    0.57 &   0.64 &  26 &
                            &    0.34 &   0.37 &  20 &
                            &    0.48 &   0.52 &  14 & 
                            &    0.48 &   0.53 &  12 &
                            &     --- &    --- & --- &
                            \\  

4861.33 &  \ion{H}{i}       &    
                            &  100.00 & 100.00 &   3 &
                            &  100.00 & 100.00 &   3 &
                            &  100.00 & 100.00 &   3 &
                            &  100.00 & 100.00 &   3 &
                            &  100.00 & 100.00 &   4 &
                            \\

4881.00 & [\ion{Fe}{iii}]   &    
                            &    0.16 &   0.16 &  45 &
                            &    0.39 &   0.38 &  45 &
                            &    0.23 &   0.23 &  50 & 
                            &    0.20 &   0.20 &  50 &
                            &     --- &    --- & --- &
                            \\  

4921.93 &  \ion{He}{i}      &    
                            &    1.28 &   1.22 &  20 &
                            &    1.16 &   1.11 &  50 &
                            &    1.25 &   1.19 &  20 & 
                            &    1.17 &   1.11 &  20 &
                            &     --- &    --- & --- &
                            \\  

4958.91 & [\ion{O}{iii}]    &    
                            &  143.30 & 132.20 &   4 &
                            &  136.50 & 126.50 &   5 &
                            &  139.00 & 128.20 &   4 & 
                            &  140.50 & 129.80 &   5 &
                            &   33.74 &  31.27 &   5 & 
                            \\

5006.84 & [\ion{O}{iii}]    &    
                            &  451.50 & 400.50 &   5 &
                            &  429.60 & 383.90 &   6 &
                            &  437.40 & 388.10 &   5 & 
                            &  440.80 & 392.00 &   5 &
                            &   95.90 &  85.74 &   5 &
                            \\

5015.68 &  \ion{He}{i}      &  
                            &    4.35 &   3.83 &  20 &
                            &    3.11 &   2.76 &  25 &
                            &    3.15 &   2.77 &  15 & 
                            &    3.10 &   2.74 &  12 &
                            &     --- &    --- & --- &
                            \\
  
5045.10 &  \ion{N}{ii}      & \multirow{2}{*}{\big\}}      
                            & \multirow{2}{*}{  0.18} & \multirow{2}{*}{   0.15} &  \multirow{2}{*}{$>$50} &
                            & \multirow{2}{*}{  0.23} & \multirow{2}{*}{   0.20} &  \multirow{2}{*}{$>$50} &
                            & \multirow{2}{*}{  0.29} & \multirow{2}{*}{   0.25} &  \multirow{2}{*}{   40} &
                            & \multirow{2}{*}{  0.23} & \multirow{2}{*}{   0.20} &  \multirow{2}{*}{   31} &
                            & \multirow{2}{*}{   ---} & \multirow{2}{*}{    ---} &  \multirow{2}{*}{  ---} &
                            \\

5047.74 &  \ion{He}{i}      \\   

5056.31 &  \ion{Si}{ii}     &  
                            &    0.16 &   0.14 & $>$50 &
                            &    0.19 &   0.16 & $>$50 &
                            &    0.43 &   0.37 &    23 & 
                            &    0.43 &   0.37 &    21 &
                            &     --- &    --- &   --- &
                           \\  

5197.90 & [\ion{N}{i}]      & \multirow{2}{*}{\big\}}  
                            & \multirow{2}{*}{  0.78} & \multirow{2}{*}{  0.59} &  \multirow{2}{*}{ 48} &
                            & \multirow{2}{*}{  0.44} & \multirow{2}{*}{  0.34} &  \multirow{2}{*}{ 42} &
                            & \multirow{2}{*}{  1.00} & \multirow{2}{*}{  0.75} &  \multirow{2}{*}{ 23} &
                            & \multirow{2}{*}{  0.61} & \multirow{2}{*}{  0.46} &  \multirow{2}{*}{ 27} &
                            & \multirow{2}{*}{  1.13} & \multirow{2}{*}{  0.82} &  \multirow{2}{*}{ 12} &
                            \\

5200.26 & [\ion{N}{i}]      \\

5270.40 & [\ion{Fe}{iii}]   &    
                            &    0.26 &   0.18 & $>$50 &
                            &    0.44 &   0.32 &    44 &
                            &    0.35 &   0.25 &    24 & 
                            &    0.31 &   0.22 &    24 &
                            &     --- &    --- &   --- &
                            \\  

5517.72 & [\ion{Cl}{iii}]   &    
                            &    1.13 &   0.65 &  13 &
                            &    1.05 &   0.62 &  10 &
                            &    1.10 &   0.63 &  10 & 
                            &    1.12 &   0.65 &  10 &
                            &     --- &    --- & --- &
                            \\  

5537.89 & [\ion{Cl}{iii}]   &    
                            &    0.81 &   0.46 &  14 &
                            &    0.97 &   0.57 &  11 &
                            &    0.84 &   0.47 &  16 & 
                            &    0.94 &   0.54 &  12 &
                            &     --- &    --- & --- &
                            \\  

5754.64 & [\ion{N}{ii}]     &    
                            &    1.03 &   0.49 &  10 &
                            &    0.80 &   0.40 &  11 &
                            &    1.07 &   0.51 &  12 & 
                            &    1.00 &   0.48 &  10 &
                            &    3.02 &   1.32 &  10 &
                            \\

5875.59 &  \ion{He}{i}      &    
                            &   33.10 &  14.48 &   4 &
                            &   30.91 &  14.23 &   4 &
                            &   32.91 &  14.44 &   4 & 
                            &   32.39 &  14.42 &   4 &
                            &   20.40 &   8.11 &   6 &
                            \\

6312.10 & [\ion{S}{iii}]    &     
                            &    4.27 &   1.40 &   5 &
                            &    4.44 &   1.56 &   4 &
                            &    4.51 &   1.49 &   4 & 
                            &    4.54 &   1.53 &   4 &
                            &     --- &    --- & --- &
                            \\  

6548.04 & [\ion{N}{ii}]     &     
                            &   28.96 &   8.28 &   5 &
                            &   32.38 &  10.00 &   3 &
                            &   30.06 &   8.64 &   3 & 
                            &   30.80 &   9.04 &   3 &
                            &  131.60 &  32.55 &  16 &
                            \\ 

6562.80 &  \ion{H}{i}       &    
                            & 1008.00 & 286.00 &   3 &
                            &  933.30 & 286.00 &   3 &
                            & 1004.00 & 286.00 &   3 & 
                            &  982.90 & 286.00 &   3 &
                            & 1167.00 & 285.80 &   5 &
                            \\   

6583.41 & [\ion{N}{ii}]     &     
                            &   91.58 &  25.68 &   3 &
                            &  106.70 &  32.35 &   3 &
                            &   92.16 &  25.96 &   3 & 
                            &   96.29 &  27.71 &   3 &
                            &  364.20 &  88.07 &   5 &
                            \\   

6678.15 &  \ion{He}{i}      &    
                            &   12.62 &   3.36 &   3 &
                            &   11.74 &   3.39 &   4 &
                            &   12.29 &   3.29 &   3 & 
                            &   11.94 &   3.35 &   3 &
                            &   11.10 &   2.53 &   5 &
                            \\

6716.44 & [\ion{S}{ii}]     &     
                            &   17.01 &   4.44 &   6 &
                            &   20.40 &   5.78 &   4 &
                            &   16.19 &   4.24 &   3 & 
                            &   18.02 &   4.83 &   3 &
                            &   85.55 &  19.09 &   8 &
                            \\    

6730.81 & [\ion{S}{ii}]     &     
                            &   17.55 &   4.54 &   5 &
                            &   20.88 &   5.88 &   3 &
                            &   17.60 &   4.58 &   3 & 
                            &   18.94 &   5.04 &   3 &
                            &   80.11 &  17.73 &  10 &
                            \\

7065.22 &  \ion{He}{i}      &    
                            &     --- &   ---  & --- &   
                            &     --- &   ---  & --- &    
                            &     --- &   ---  & --- &    
                            &     --- &   ---  & --- &    
                            &   12.03 &   2.21 &   5 &
                            \\

7135.80 & [\ion{Ar}{iii}]   &    
                            &     --- &   ---  & --- &   
                            &     --- &   ---  & --- &    
                            &     --- &   ---  & --- &    
                            &     --- &   ---  & --- &    
                            &   46.84 &   8.27 &   8 &
                            \\
                            
\noalign{\smallskip}
c(H$\beta$) & \multicolumn{3}{l}{H$\alpha/$H$\beta$} 
            & 1.65 & & & & 1.55 & & & & 1.65 & & & & 1.62 & & & & 1.84 \\
c(H$\beta$) & \multicolumn{3}{l}{H$\gamma/$H$\beta$, H$\delta/$H$\beta$} 
            & 1.42 & & & & 1.12 & & & & 0.91 & & & & 1.20 & & & & 1.53 \\
                                                       
\noalign{\smallskip}
\hline \\[2cm]
\end{tabular}                          
\end{table*}

\section{Chemical abundance analysis}
\label{abundance analysis}
The emission line intensities listed in Table~\ref{line intensities} were used 
for chemical abundance analysis of the nebulae.

\subsection{Electron densities and temperatures}
\label{electron densities and temperatures}

The electron temperature estimates referred as $T_{{\rm e}}(\ion{O}{iii})$ and 
$T_{{\rm e}}(\ion{N}{ii})$ were derived from the $[\ion{O}{iii}](\lambda\,4959 +
\lambda\,5007)/\lambda\,4363$ and $[\ion{N}{ii}](\lambda\,6548 + 
\lambda\,6583)/\lambda\,5755$ line intensity ratios and the electron density 
estimates $N_\mathrm{e}(\ion{S}{ii})$ and $N_\mathrm{e}(\ion{Cl}{iii})$ from the  
$[\ion{S}{ii}]\lambda\,6716/\lambda\,6731$ and 
$[\ion{Cl}{iii}]\lambda\,5517/\lambda\,5537$ ratios, respectively. These 
electron temperatures and densities were obtained by solving numerically the 
equilibrium equations for an $n$-level atom ($5 \leq n \leq 9$) using the {\it 
temden} routine of the {\it nebular} package of the {\it STSDAS/IRAF}, using the 
same atomic parameters as in \citet{Krabbe & Copetti 2005}. Table~\ref{table 
densities and temperatures} lists the electron densities and temperatures 
obtained. For $N_\mathrm{e}(\ion{Cl}{iii})$ only upper limits were presented 
because the error interval for the $[\ion{Cl}{iii}]\lambda\,5517/\lambda\,5537$ 
ratio extends beyond the low density limit. 
%
%
\begin{table*}
\caption{Electron densities and temperatures (in units of cm$^{-3}$ and K, respectively)}
\label{table densities and temperatures}
\begin{tabular}{lrrrrr}
\hline
\noalign{\smallskip}
& NGC\,2579, A  & 
NGC\,2579, B  & 
NGC\,2579, C & 
NGC\,2579, ABC &
ESO\,370-9 \\
\hline
\noalign{\smallskip}
$N_\mathrm{e}$ (\ion{S}{ii})   &      699 $\pm$ 219 &    641  $\pm$ 126  &     869 $\pm$ 129 &     745 $\pm$ 114  &     443 $\pm$ 262 \\ 
$N_\mathrm{e}$ (\ion{Cl}{iii}) &         $<$ 1\,100 &         $<$ 2\,900 &        $<$ 1\,600 &         $<$ 2\,100 &                   \\
$T_\mathrm{e}$ (\ion{N}{ii})   &  11\,550 $\pm$ 620 &  9\,630 $\pm$ 430  & 11\,670 $\pm$ 680 & 11\,030 $\pm$ 520  & 10\,300 $\pm$ 790 \\ 
$T_\mathrm{e}$ (\ion{O}{iii})  &   9\,490 $\pm$ 230 &  9\,140 $\pm$ 240  &  8\,620 $\pm$ 190 &  8\,960 $\pm$ 210  &                   \\
\noalign{\smallskip}
\hline
\end{tabular}
\end{table*}

\subsection{Abundance determination}
\label{abundance determination}

We have derived ionic and total abundances of He, N, O, Ne, S, Cl, and Ar. 
Although we have detected some recombination lines of heavy elements, their 
intensities are too uncertain to be used in the abundance analysis. We thus  
relied only upon the collisionally excited lines to calculate abundances of 
metals. Based on the similarity of the ionization potentials, we adopted 
for NGC\,2579 the electron temperature $T_{{\rm e}}(\ion{N}{ii})$ for 
the N$^{+}$, O$^{+}$, and S$^{+}$ ionic zones, and $T_{{\rm e}}(\ion{O}{iii})$ 
for O$^{+*}$, Ne$^{++}$, S$^{++}$ and Cl$^{++}$. For ESO\,370-9, because we do 
not have measured $T_{{\rm e}}(\ion{O}{iii})$, we assumed an electron 
temperature 20\% lower than $T_{{\rm e}}(\ion{N}{ii})$ for the double ionized 
ions. Since the abundance estimates are barely dependent on the assumed electron 
density, we adopted for all ionic zones fixed densities of 750 and 440 
cm$^{-3}$ (the [\ion{S}{ii}] densities from the integrated spectra) for 
NGC\,2579 and ESO\,370-9, respectively. Ionic metal abundances were obtained 
with the {\it ionic} routine of the {\it nebular} package of the {\it 
STSDAS/IRAF}. The references for atomic parameters used are listed in Table~2 of 
\citet{Krabbe & Copetti 2006}. The ionic helium abundances were derived from the 
strongest lines $\lambda4471$, $\lambda5876$, and $\lambda6678$, using the 
\ion{He}{i} emissivities of \citet{Benjamin et al. 1999}, which are corrected by 
the effects of collisional excitation.

The total abundance is the sum of all measured ionic abundances of a given 
element corrected for unseen ionization stages. We have followed an ionization 
correction scheme adapted from that of \citet{Liu et al. 2000} for Cl, and of 
\cite{Kingsburgh & Barlow 1994} for the other elements based on the similarities 
between the ionization potentials of different ions and on photoionization 
models. No precise abundance measurement of neutral elements was possible for 
NGC\,2579. The detected lines of [\ion{N}{i}] were very weak and the 
[\ion{O}{i}] lines were heavily contaminated by telluric emission. Assuming that 
the fraction of any neutral heavy elements ${\rm X^0/X}$ is the same as that of 
neutral hydrogen ${\rm H^0/H}$, which is an adequate assumption according to 
\citet{Kingsburgh & Barlow 1994}, especially for the elements O and N, we can 
ignore the abundances of all neutral species, since in this case we have, as a 
very good approximation, \begin{equation}
{\rm X/H \approx \sum_{i = 1}^\infty X^{i+}/H^+}.
\end{equation}
Based on the spectral types of the ionizing stars of NGC\,2579 (see 
Sect.~\ref{ionizing stars}), we do not expect a significant amount of neutral 
gas inside the nebula. We have verified that the helium lines present uniform 
intensities relative to H$\beta$ across the nebula, indicating that this element 
is fully ionized, as expected. This is not the case of ESO\,370-9, which is a 
low ionization nebula. On the other hand, we have not detected any high 
excitation emission line, as the \ion{He}{ii} and [\ion{Ar}{iv}] lines, implying 
that triple ionized ions are not present in significant quantities. These and 
other similar arguments justify the following expressions adopted for 
calculating the total abundances:
\begin{equation}
{\rm O/H = O^+/H^+ + O^{++}/H^+},
\end{equation}

\begin{equation}
{\rm N/H = (1+O^{++}/O^+) \times N^+/H^+},
\end{equation}

\begin{equation}
{\rm Ne/H = (1+O^+/O^{++}) \times Ne^{++}/H^+},
\end{equation}

\begin{equation}
{\rm Ar/H = 1.87 \times Ar^{++}/H^+},
\end{equation}

\begin{equation}
{\rm S/H = (1-(1+O^+/O^{++})^{-3})^{-1/3} \times (S^+/H^+ + S^{++}/H^+)},
\end{equation}

\begin{equation}
{\rm Cl/H = (S/H)/(S^{++}/H^+) \times Cl^{++}/H^+}.
\end{equation}
To derive the total abundances of ESO\,370-9, in the absence of measurements of 
[\ion{O}{ii}] and [\ion{S}{iii}] lines, we have relied upon the abundance ratios 
He$^+$/He = 46.1\%, O$^{++}$/O = 24.9\%, Ar$^{++}$/Ar = 41.6\%, and S$^+$/S = 
15.3\%, obtained by fitting the intensities of the strong emission lines with 
the photoionization code Cloudy \citep{Ferland et al. 1998}. 
%
%
\begin{table}
\caption{Ionic and total abundances (in the scale ${\rm 12+\log(X/H)}$)}
\label{abundances}
\begin{tabular}{lllllllll}
\hline
\noalign{\smallskip}
$\lambda_0$ & Abund. & NGC\,2579 & ESO\,379-9 & Orion \\
\hline
\noalign{\smallskip}
4471 & He$^+$/H$^+$    & 10.93 $\pm$ 0.01 & 10.83 $\pm$ 0.06 &       \\
5876 & He$^+$/H$^+$    & 11.02 $\pm$ 0.02 & 10.77 $\pm$ 0.03 &       \\
6678 & He$^+$/H$^+$    & 10.94 $\pm$ 0.01 & 10.83 $\pm$ 0.02 &       \\
avg. & He$^+$/H$^+$    & 10.96 $\pm$ 0.04 & 10.81 $\pm$ 0.02 &       \\
     & He/H            & 10.96 $\pm$ 0.04 & 11.14 $\pm$ 0.18 & 10.99 \\
3727 & O$^+$/H$^+$     &  7.65 $\pm$ 0.16 &                  &       \\
4959 & O$^{++}$/H$^+$  &  8.32 $\pm$ 0.06 &  7.75 $\pm$ 0.11 &       \\
5007 & O$^{++}$/H$^+$  &  8.30 $\pm$ 0.05 &  7.73 $\pm$ 0.11 &       \\
avg. & O$^{++}$/H$^+$  &  8.31 $\pm$ 0.06 &  7.73 $\pm$ 0.11 &       \\
     & O/H             &  8.39 $\pm$ 0.05 &  8.33 $\pm$ 0.20 &  8.47 \\
5518 & Cl$^{++}$/H$^+$ &  5.02 $\pm$ 0.05 &                  &       \\
5538 & Cl$^{++}$/H$^+$ &  5.04 $\pm$ 0.05 &                  &       \\
avg. & Cl$^{++}$/H$^+$ &  5.03 $\pm$ 0.05 &                  &       \\
     & Cl/H            &  5.16 $\pm$ 0.08 &                  &  5.16 \\
3869 & Ne$^{++}$/H$^+$ &  7.60 $\pm$ 0.08 &                  &       \\
3968 & Ne$^{++}$/H$^+$ &  7.54 $\pm$ 0.11 &                  &       \\
avg. & Ne$^{++}$/H$^+$ &  7.59 $\pm$ 0.08 &                  &       \\
     & Ne/H            &  7.68 $\pm$ 0.06 &                  &  7.71 \\
6548 & N$^+$/H$^+$     &  6.61 $\pm$ 0.12 &  7.24 $\pm$ 0.10 &       \\
6584 & N$^+$/H$^+$     &  6.63 $\pm$ 0.12 &  7.21 $\pm$ 0.08 &       \\
avg. & N$^+$/H$^+$     &  6.63 $\pm$ 0.12 &  7.22 $\pm$ 0.09 &       \\
     & N/H             &  7.37 $\pm$ 0.06 &  7.35 $\pm$ 0.12 &  7.62 \\
4069 & S$^+$/H$^+$     &  5.18 $\pm$ 0.12 &       \\
6716 & S$^+$/H$^+$     &  5.31 $\pm$ 0.13 &  5.93 $\pm$ 0.08 &       \\
6731 & S$^+$/H$^+$     &  5.31 $\pm$ 0.12 &  5.93 $\pm$ 0.08 &       \\
avg. & S$^+$/H$^+$     &  5.30 $\pm$ 0.12 &  5.93 $\pm$ 0.08 &       \\
6312 & S$^{++}$/H$^+$  &  6.78 $\pm$ 0.08 &                  &       \\
     & S/H             &  6.91 $\pm$ 0.11 &  6.75 $\pm$ 0.19 &  6.97 \\
7136 & Ar$^{++}$/H$^+$ &                  &  6.05 $\pm$ 0.08 &       \\
     & Ar/H            &                  &  6.43 $\pm$ 0.19 &  6.52 \\
\noalign{\smallskip}                        
\hline                                      
\end{tabular}                               
\end{table}                                 
                                            
Table~\ref{abundances} presents the ionic abundances obtained from different 
lines, the weighted-by-the-line-intensity average values, and the total 
abundances for the sum of the spectra obtained for NGC\,2579 at positions A, B, 
and C, and for ESO\,370-9, all expressed in the logarithmic scale 
${\rm 12+\log(X/H)}$. The error estimates for NGC\,2579 correspond to the mean 
absolute deviation of the abundances obtained at these three different 
positions. For ESO\,370-9 the errors in the ionic abundances were obtained by 
propagation of the errors in the line intensities and electron temperature, 
while in the case of total abundances, an error of 50\% in the ionization 
correction factors was assumed. For comparison, we also show in this table the 
total abundances obtained by \citet{Esteban et al. 1998} for the Orion Nebula. 
No temperature fluctuation correction was applied to the total abundances. 
Although the total abundance estimates for ESO\,370-9 required large and 
uncertain ionization corrections they are quite similar to those of NGC\,2579.

\section{Electron density and radial velocity maps}
\label{electron density and velocity maps}

\begin{figure}
\centering
\resizebox{\hsize}{!}{\includegraphics*[angle=-90]{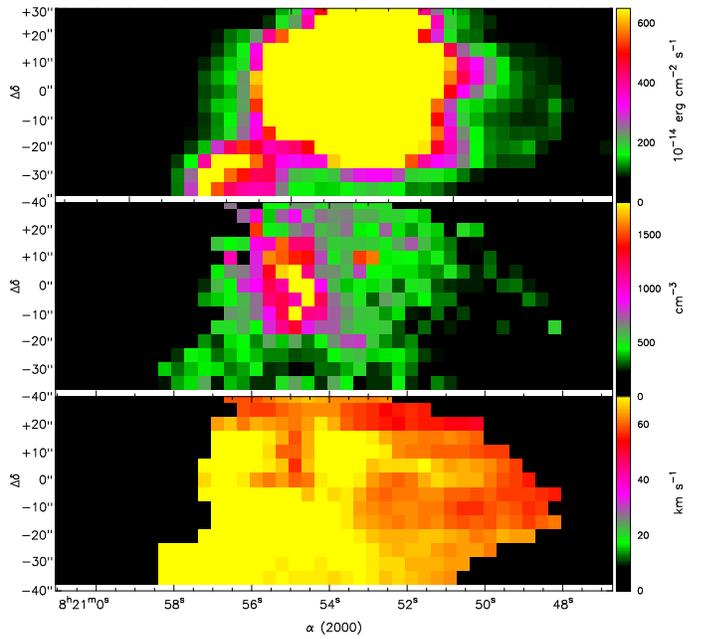}}
\caption{Maps of the H$\alpha$ flux (upper panel), electron density (mid panel), 
and LSR radial velocity (lower panel)} 
\label{maps}
\end{figure}

Figure~\ref{maps} presents the $5\arcsec \times 5\arcsec$ resolution maps of 
H$\alpha$ flux, $[\ion{S}{ii}]$ electron density and radial velocity with 
respect to the local standard of rest constructed from the equally spaced 
long-slit spectroscopic observations. A mask, defined by the positions with 
H$\alpha$ flux larger than 0.1\% of the peak value, was used to delimit the 
nebula. Only about half of ESO\,370-9 was covered by the observations. 
Table~\ref{statistics} presents some statistics of density and velocity 
measurements, including the number $N$ of distinct nebular areas, the median, 
the first and third quartiles, $Q1$ and $Q3$ respectively (limits between which 
50\% of the values lie), the minimum and maximum, and the weighted by the 
H$\alpha$ flux mean and the standard deviation $\sigma$.

NGC\,2579 and ESO\,370-9 sit on a common velocity plateau of $v_\mathrm{LSR}$ 
$\approx$ 60$-$66 km s$^{-1}$, indicating that both objects are at a similar 
distance. From CO observations towards the area, a comparable velocity of 
$v_\mathrm{LSR}$ = 68.3$\pm$0.3 km s$^{-1}$ was obtained by \citet{Brand et al. 
1987}. Parts of NGC\,2579, especially in the western and in the northern-central 
areas, show significantly lower velocities, indicating an internal systematic 
flow of ionized matter, with the gas streaming away from the main body of the 
nebula and from the associated molecular cloud with velocities of $\approx$ 15 
km s$^{-1}$.

As previously indicated by the spatial profile of electron density along a 
single direction across the nebula (corresponding to offset declination 
$\Delta\delta = 0$) obtained by \citet{Copetti et al. 2000}, NGC\,2579 presents 
a non-uniform density structure. The electron density presents steep gradients, 
especially towards the east, with the density ranging from about 1800 cm$^{-3}$ 
at the brightest eastern-central areas to less than 100 cm$^{-3}$ at the outer 
parts of the nebula. Both the velocity and density structures of NGC\,2579 
suggest that a `blister' \citep{Israel 1978} or `champagne' \citep{Tenorio-Tagle 
1979} flow is taking place in the nebula.
%
%

%
%
\begin{table}
\caption{Electron density and radial velocity statistics}
\label{statistics}
\begin{tabular}{l@{~~~~~}l@{~~~~~}l@{~~~~~}l@{~~~~~}l@{~~~~~}l}
\hline
\noalign{\smallskip}
& \multicolumn{2}{c}{$v_\mathrm{LSR}$ (km s$^{-1}$)} & & \multicolumn{2}{c}{$N_\mathrm{e}$ (cm$^{-3}$)}\\ 
\noalign{\smallskip}
\cline{2-3}\cline{5-6}
\noalign{\smallskip}
& NGC\,2579 & ESO\,379-9 & & NGC\,2579 & ESO\,379-9 \\
\noalign{\smallskip}
\hline
\noalign{\smallskip}
$N$      & 298   & 40    & & 298    & 40  \\
min      & 42.22 & 56.76 & & $<$100 & 133 \\
$Q1$     & 50.00 & 58.46 & & 225    & 225 \\
median   & 53.03 & 61.33 & & 376    & 303 \\
$Q3$     & 57.00 & 63.91 & & 518    & 354 \\
max.     & 68.98 & 66.82 & & 1886   & 530 \\
mean     & 54.70 & 61.40 & & 435    & 305 \\
$\sigma$ & 5.28  & 2.68  & & 491    & 76  \\
\noalign{\smallskip}
\hline
\end{tabular}
\end{table}

\section{Ionizing stars}
\label{ionizing stars}

Table~\ref{table Studied stars} presents the designations and equatorial 
coordinates of the five brightest stars towards NGC\,2579 and ESO\,370-9 which 
were the targets for our spectroscopic and photometric studies for the 
identification of the ionizing stars of these two nebulae and for the 
determination of their distances. Hereafter we will designate these stars by 
their entry numbers in this table.

%
%
\begin{table*}
\caption{Studied stars}
\label{table Studied stars}
\begin{tabular}{llllllll}
\hline
\noalign{\smallskip}
\multicolumn{4}{l}{Star designation} & & \multicolumn{2}{l}{Position (J2000)}\\ 
\noalign{\smallskip}
\cline{1-4}\cline{6-7}
\noalign{\smallskip}
$N$ & 
VdBH$^\dagger$ & 
DENIS &
Other & & 
$\alpha$ &
$\delta$ & 
Nebula \\ 
\noalign{\smallskip}
\hline
\noalign{\smallskip}
1 & 13a A & J082052.9-361251 & CD-35\, 4502 & & 08 20 52.92 & -36 12 51.1 & NGC\,2579   \\
2 & 13a B & J082052.8-361258 &              & & 08 20 52.83 & -36 12 58.0 & NGC\,2579  \\
3 & 13b A & J082054.8-361258 &              & & 08 20 54.86 & -36 12 58.9 & NGC\,2579  \\
4 & 13b B & J082055.0-361306 &              & & 08 20 55.05 & -36 13 06.0 & NGC\,2579  \\
5 & 13c   & J082056.9-361342 &              & & 08 20 56.97 & -36 13 42.5 & ESO\,370-9 \\
\noalign{\smallskip}
\hline
\noalign{\smallskip}
\end{tabular}
\\
{\scriptsize
$\dagger$Original designations of VdBH 13a and VdBH 13b broken in two stars by
\citet{Rousseau & Perie 1996} and in this paper, respectively.} 

\end{table*}

%
%
\begin{table*}
\caption{$UBV$ photometry and properties of the studied stars}
\label{table UBV photometry}
\begin{tabular}{lrrrlrrccll}
\hline
\noalign{\smallskip}
Star & 
$V$~~~~~~~~~~~&
$U-V$~~~~~~ &  
$B-V$~~~~~~ &
Sp &
$M_V$ &
$(B-V)_0$ &
$E(B-V)$ & 
$D$ (kpc) &
$Q({\rm H}^0)$ (s$^{-1}$) \\ 
\noalign{\smallskip}
\hline
\noalign{\smallskip}
1 & 10.385 $\pm$ 0.008 &    1.377 $\pm$ 0.024 & 1.553 $\pm$ 0.017 & K1 III &    0.61 &    1.07~~~ & 0.48 & 0.45 &       \\
2 & 13.055 $\pm$ 0.015 & $-$0.104 $\pm$ 0.036 & 0.934 $\pm$ 0.029 & O5 V   & $-$5.33 & $-$0.33~~~ & 1.26 & 7.82 & 49.48 \\
3 & 13.607 $\pm$ 0.012 & $-$0.237 $\pm$ 0.036 & 1.038 $\pm$ 0.026 & O6.5 V & $-$4.99 & $-$0.32~~~ & 1.36 & 7.54 & 49.17 \\
4 & 13.910 $\pm$ 0.034 & $-$0.099 $\pm$ 0.084 & 1.028 $\pm$ 0.072 & O8 V   & $-$4.66 & $-$0.32~~~ & 1.35 & 7.56 & 48.80 \\
5 & 13.748 $\pm$ 0.004 & $-$0.184 $\pm$ 0.019 & 0.958 $\pm$ 0.008 & O8.5 V & $-$4.55 & $-$0.31~~~ & 1.27 & 7.47 & 48.64 \\
\noalign{\smallskip}
\hline
\end{tabular}
\end{table*}
 
\subsection{Spectral classification}
 
Star 1, which is the brightest star, was the only one with a previous spectral 
classification. We confirm the spectral type of K1\,III attributed for this star 
by \citet{Herbst 1975}. For the spectra of stars 2, 3 and 4  embedded in bright 
parts of NGC\,2579 the background subtraction to eliminate nebular emission 
lines was difficult, so a spectral classification based solely on Balmer or 
\ion{He}{i} lines would be unreliable. Fortunately, these stars show relatively 
strong \ion{He}{ii} absorption lines, which could not possibly be a fake result 
of the background subtraction. So, from a visual inspection the spectral atlas 
of O stars by \citet{Walborn & Fitzpatrick 1990} and especially from the 
comparison of the equivalent widths of the \ion{He}{ii} absorption lines with 
those from \citet{Conti 1973} and \citet{Conti & Alschuler 1971} we attribute 
spectral type of O5\,V, O6.5\,V, and O8\,V for stars 2, 3, and 4, respectively. 
For star 5, the central star in ESO\,370-9, we found a spectral type of O8.5\,V.

\subsection{Distance}
\label{distance}

Table~\ref{table UBV photometry} presents the results of the $UBV$ photometry 
for the five stars studied, together with the spectral type Sp and some physical 
properties derived from the spectral classification, namely the visual absolute 
magnitude $M_V$ for O stars taken from \citet{Vacca et al. 1996}, the intrinsic 
color $(B-V)_0$ (and $M_V$ for the K1 III star) from \citet{Schmidt-Kaler 
1982}, and the Lyman continuum photon flux $Q({\rm H}^0)$ from \citet{Schaerer & 
de Koter 1997}. Also shown in this table are the color excess $E(B-V) = (B-V) - 
(B-V)_0$ and the heliocentric distance $D$ calculated assuming a visual 
reddening of $A_V = 3.1 E(B-V)$. It is clear from this table that star 1 is a 
cold foreground star. The others, including star 5 in ESO\,370-9, are hot O type 
stars with comparable reddening and distance estimates. Thus, we conclude that 
NGC\,2579 and ESO\,370-9 are at a similar heliocentric distance of 7.6 kpc. The 
dispersion among the distance estimates for these four stars is only 2\%, and 
the propagated photometric error is quite small. However, the photometric distance is strongly dependent on the assumed properties of the 
stars, especially the value of $M_V$. The adoption of the absolute visual 
magnitudes for O stars from \citet{Schmidt-Kaler 1982} would result in 13\% 
higher distances, while based on the recent calibration of \citet{Martins et al. 
2005} we would obtain distances 12\% lower. On the other hand, the use of 
$(B-V)_0$ from other sources, for example from \citet{Fitzgerald 1970}, would 
produce distances only 1$-$2\% different. Thus, taking into account the 
uncertainties in the intrinsic stellar properties adopted, we estimate an error 
for the photometric distance of about 0.9 kpc

\citet{Brand & Blitz 1993} found for the object BBW\,138 ($\equiv$ Bran\,138), 
positively identified with NGC\,2579, a photometric distance of 11.43$\pm$2.33 
kpc, which is 50\% higher than our distance estimate. They used the Walraven 
photometric system ($VBLUW$) to derive the distance, as described in 
\citet{Brand & Wouterloot 1988}, but the specific photometric data for 
NGC\,2579 has not been published (as far as we know). Therefore, we refrain 
from making deeper comparison between our results.  

The radial velocity map (Fig.~\ref{maps}) clearly indicates that NGC\,2579 and 
ESO\,370-9 are at a similar distance. From the measurements of the radial 
velocity $v_\mathrm{LSR}$ (CO) = 68.3$\pm$0.3 km s$^{-1}$, from CO observations 
obtained by \citet{Brand et al. 1987} (for BBW\,138), $v_\mathrm{LSR}$ (5 GHz) = 
64 km s$^{-1}$, from radio recombination lines H 109$\alpha$ and H 110$\alpha$ 
by \citet{Caswell & Haynes 1987} (for the object refereed by 254.676 +0.229), 
and $v_\mathrm{LSR}$ (H$\alpha$) = 63 km s$^{-1}$, the mean velocity of the 
velocity plateau from the present papers (see Sect.~\ref{electron density and 
velocity maps}), we have calculated a kinematic distance of 7.4$\pm$1.4 kpc for 
NGC\,2579 by means of the rotation curve for the Galaxy from \citet{Brand & 
Blitz 1993} assuming the solar galactocentric distance $R_\odot$ = 8.5 kpc and 
velocity $\Theta_\odot$ = 220 km s$^{-1}$. This kinematic distance is entirely 
compatible with the photometric distance estimated by us. 

The distance of the nebula can also be estimated from the observed fluxes in 
Balmer lines assuming no leakage of ionizing photons from
\begin{equation}
D = \sqrt{\frac{ 
h\nu_{\mathrm{H}\beta} \,
\alpha^\mathrm{eff}_{\mathrm{H}\beta}(\mathrm{H}^0,T_\mathrm{e}) \, 
Q(\mathrm{H}^0)
}{
4\pi \,
\alpha_\mathrm{B}(\mathrm{H}^0,T_\mathrm{e}) \,
F(\mathrm{H}\beta) \,
10^{c(\mathrm{H}\beta)
}}},
\end{equation}
where $h\nu_{\mathrm{H}\beta}$ is the energy of the H$\beta$ photon, 
$\alpha^\mathrm{eff}_{\mathrm{H}\beta}(\mathrm{H}^0,T_\mathrm{e})$ is the 
effective recombination coefficient for H$\beta$ and 
$\alpha_\mathrm{B}(\mathrm{H}^0,T_\mathrm{e})$ is the total recombination 
coefficient to excited levels of H. Comparing the integrated H$\alpha$ flux 
obtained in this paper (see Sect.~\ref{Halpha photometry}) with the H$\beta$ 
flux of $\log F(\mathrm{H}\beta) = -10.72$ (in units of erg cm$^{-2}$ s$^{-1}$) 
from \citet{Copetti 2000}, we calculate a global extinction of $c$(H$\beta$) = 
1.91 (from the stellar photometry we get $c(\mathrm{H}\beta) = 1.5 E(B-V) = 
1.96$, and from the spectroscopy at selected positions $c(\mathrm{H}\beta) 
\approx 1.6$). The total rate of ionizing photons $Q(\mathrm{H}^0)$ was obtained 
adding together the values shown in Table~\ref{table UBV photometry} for 
individual stars. Although we have to make strong assumptions in this method, 
we obtained a comparable distance of 10.5 kpc, only 40\% higher than the 
photometric distance. This distance estimate is also strongly dependent on the 
adopted parameters for O stars, and the uncertainties on $Q(\mathrm{H}^0)$ are 
high. For example, based on the models of \citet{Martins et al. 2005} we would 
derive about half of the ionizing photon flux obtained from the calibration with 
the spectral type by \citet{Schaerer & de Koter 1997}. More importantly, 
inverting the arguments, we can show that, assuming the photometric distance 
obtained for the pair NGC\,2579 and ESO\,370-9, stars 2, 3, 4, and 5 may be 
solely responsible for the required budget of ionizing photons. For a distance 
of 7.6 kpc (and $c$(H$\beta$) = 1.91), from the observed H$\alpha$ fluxes we 
derive $\log Q(\mathrm{H}^0) = 49.40$ for NGC\,2579 and $\log Q(\mathrm{H}^0) = 
48.46$ for ESO\,370-9 (in units of s$^{-1}$), while from the spectral types we 
estimate $\log Q(\mathrm{H}^0) = 49.71$ and 48.64, respectively. These 
figures may be considered compatible, since errors of the order of 0.30 dex are 
expected for $Q(\mathrm{H}^0)$.

\section{Discussion}

NGC\,2579 is an interesting object not only due to its high surface brightness 
but especially because of its location in the Galaxy. The estimated heliocentric 
distance of $D = 7.6$ kpc corresponds to a galactocentric distance of $R = 12.8$ 
kpc. With a moderate interstellar extinction of about 0.5 magnitudes in $V$ per 
kiloparsec, NGC\,2579 may contribute significantly to the studies of the 
abundance gradients in the outer Galaxy, since it has been very difficult to 
find in this part of the Galaxy objects bright enough to allow direct abundance 
determinations \citep{Fich & Silkey 1991, Vilchez & Esteban 1996}. In fact, 
NGC\,2579 is one of the most distant Galactic \ion{H}{ii} region for which the 
emission line ratio $[\ion{O}{iii}](\lambda\,4959 + 
\lambda\,5007)/\lambda\,4363$ has been already measured. The metal abundances  
measured in NGC\,2579 are slightly lower (by 24\% in mean and by 20\% for the 
O/H) than those in the Orion Nebula. The helium abundance is lower than in Orion 
by 7\%. From the comparison of the oxygen abundances of these two nebulae alone 
we would derive a shallow abundance gradient of $-0.02 \pm 0.01$ dex/kpc for the 
galactocentric distance range $8.8 < R~(\mathrm{kpc}) < 12.8$. A more complete 
investigation of the impact of the chemical composition determination of 
NGC\,2579 on the abundance gradients in the Galaxy will be present elsewhere.

\subsection{The nature of ESO\,370-9}
\label{nature of eso 370-9}

ESO\,370-9 is a roughly elliptical 40\arcsec$\times$50\arcsec\ ringed nebula 
with a star in the middle. Because of this morphology it has been misclassified 
as a planetary nebula. With a mean linear diameter of $\approx$ 1.6 pc it is 
comparable in size with the largest planetary nebulae. However, ESO\,370-9 is
definitively too massive to be a planetary nebula. Assuming an electron 
density of $N_\mathrm{e}$ = 440 cm$^{-3}$ (the integrated [\ion{S}{ii}] density 
from Table \ref{table densities and temperatures}), an electron temperature of 
$T_\mathrm{e} = 10^4$ K, a Lyman continuum photon flux of $\log Q(\mathrm{H}^0) 
= 48.46$ (in units of s$^{-1}$), an ionized helium abundance of $y^+ = 0.05$ 
and a total helium abundance of $y = 0.1$, we estimate a mass of $M$ $\approx$ 
28 $M_\odot$ using the expression 
\begin{equation}
M = \frac{m_\mathrm{p} (1+4y)}{(1+y^+) \, 
\alpha_\mathrm{B}(\mathrm{H}^0,T_\mathrm{e}) } 
\frac{Q(\mathrm{H}^0)}{N_\mathrm{e}},
\end{equation}
where $m_\mathrm{p}$ is the proton mass. Besides, the chemical abundances of 
ESO\,370-9, especially the relative abundance of nitrogen to oxygen of N/O 
$\approx 0.10$, are more typical of \ion{H}{ii} regions. Planetary nebulae 
usually present higher N/O abundance ratios, on average by a factor of 5 and in 
extreme cases by a factor larger than 10 \citep{Peimbert & Torres-Peimbert 1971, 
Kingsburgh & Barlow 1994, Perinotto et al. 2004}. So, we conclude that 
ESO\,370-9 is a small and relatively low excitation \ion{H}{ii} region ionized 
by a single O8.5\,V star located at a distance similar to that of NGC\,2579 
($\approx$ 7.6 kpc), which leaves the possibility that these two objects are 
physically associated.

%
%
\section{Conclusions}

We have presented the first comprehensive optical observational study on the 
nebular and stellar properties of the Galactic \ion{H}{ii} regions NGC\,2579 and 
ESO\,370-9, which includes the determination of electron temperature and density, 
chemical composition, and the study of the density and radial velocity 
structures of the nebulae. We have also pursued the identification and spectral 
classification of the ionizing stars, and the determination of their distances. 
The nature of ESO\,370-9, usually misclassified as planetary or reflection 
nebula, is discussed. The main conclusions are:

\begin{enumerate}

\item
The chemical abundances of He, N, O, Ne, S, and Cl measured in NGC\,2579 are 
slightly lower than those in the Orion Nebula, the metal abundances by about 
24\% on average and the helium abundance by 7\%, which is consistent with the 
chemical composition gradient in the Galaxy.

\item
NGC\,2579 is ionized by three O stars of spectral types O5\,V, O6.5\,V, and 
O8\,V, while ESO\,370-9 is ionized by a single O8.5\,V star. These stars are 
entirely capable of being solely responsible for the required ionizing photon 
fluxes estimated from Balmer recombination lines, although other cooler stars 
should be present.

\item
NGC\,2579 and ESO\,370-9 are at a similar distance. We have estimated from 
spectroscopic parallax a heliocentric distance of 7.6 $\pm$ 0.9 kpc for both 
objects, which corresponds to a galactocentric distance of 12.8 $\pm$ 0.7 kpc 
(for $R_\odot$ = 8.5 kpc). A similar kinematic distance of 7.4 $\pm$ 1.4 kpc was 
derived from the H$\alpha$ velocity field.

\item
NGC\,2579 presents a steep density gradient, with the electron density ranging 
from about 1800 cm$^{-3}$ at the brightest eastern-central areas to less than 100 
cm$^{-3}$ at the outer parts. Both the velocity and density structures of 
NGC\,2579 suggest that a `blister' or `champagne' flow is taking place in the 
nebula.

\item
The estimated mass of gas of $\approx$ 28 $M_\odot$ for ESO\,370-9 indicates 
that it can not be a planetary nebula. With a diameter of $\approx$ 
1.6 pc, ESO\,370-9 is a small and relatively low excitation \ion{H}{ii} 
region ionized by a single O8.5\,V star. It is located at about the same 
distance as NGC\,2579.

\item
The Galactic \ion{H}{ii} region \object{NGC\,2579} has been neglected for a long 
time due to identification problems which persisted until recently. It has been 
misclassified as planetary or reflection nebula and confused with other objects. 
Because of its high surface brightness, angular size of few arcminutes and 
relatively low interstellar extinction, it is an ideal object for investigations 
in the optical range. Besides this, its location at the outer Galaxy and its 
high excitation make NGC\,2579 an essential object for the studies of the 
Galactic chemical abundance gradients.

\end{enumerate}

\begin{acknowledgements}
This work was partially supported by the Brazilian institutions CAPES, CNPq and 
FAPERGS. We thank the referee, Manuel Peimbert, for helpful comments and suggestions.
\end{acknowledgements}


\begin{thebibliography}{}

\bibitem[Acker et al.(1992)]{Acker et al. 1992} Acker, A., Marcout, J., 
Ochsenbein, F., Stenholm, B., \& Tylenda, R.\ 1992, Garching: European 
Southern Observatory, 1992,  

\bibitem[Archinal \& Hynes(2003)]{Archinal & Hynes 2003} Archinal, B.~A., \& 
Hynes, S.~J.\ 2003, Star clusters, Willmann-Bell, 2003.,  

\bibitem[Benjamin et al.(1999)]{Benjamin et al. 1999} Benjamin, R.~A., 
Skillman, E.~D., \& Smits, D.~P.\ 1999, \apj, 514, 307 

\bibitem[Brand et al.(1986)]{Brand et al. 1986} Brand, J., Blitz, L., \& 
Wouterloot, J.~G.~A.\ 1986, \aaps, 65, 537 

\bibitem[Brand et al.(1987)]{Brand et al. 1987} Brand, J., Blitz, L., 
Wouterloot, J.~G.~A., \& Kerr, F.~J.\ 1987, \aaps, 68, 1 

\bibitem[Brand \& Blitz(1993)]{Brand & Blitz 1993} Brand, J., \& Blitz, 
L.\ 1993, \aap, 275, 67 

\bibitem[Brand \& Wouterloot(1988)]{Brand & Wouterloot 1988} Brand, J., \& 
Wouterloot, J.~G.~A.\ 1988, \aaps, 75, 117 

\bibitem[Caswell \& Haynes(1987)]{Caswell & Haynes 1987} Caswell, J.~L., \& 
Haynes, R.~F.\ 1987, \aap, 171, 261 

\bibitem[Conti(1973)]{Conti 1973} Conti, P.~S.\ 1973, \apj, 179, 
161 

\bibitem[Conti \& Alschuler(1971)]{Conti & Alschuler 1971} Conti, P.~S., \& 
Alschuler, W.~R.\ 1971, \apj, 170, 325 

\bibitem[Copetti(2000)]{Copetti 2000} Copetti, M.~V.~F.\ 2000, \aaps, 147, 93 

\bibitem[Copetti \& Dottori(1989)]{Copetti & Dottori 1989} Copetti, M.~V.~F., 
\& Dottori, H.~A.\ 1989, \aaps, 77, 327 

\bibitem[Copetti et al.(2000)]{Copetti et al. 2000} Copetti, M.~V.~F., 
Mallmann, J.~A.~H., Schmidt, A.~A., \& Casta{\~n}eda, H.~O.\ 2000, \aap, 
357, 621 

\bibitem[Dreyer(1888)]{Dreyer 1888} Dreyer, J.~L.~E.\ 1888, New General 
Catalogue, Royal Astron. Soc., London 

\bibitem[Esteban et al.(1998)]{Esteban et al. 1998} Esteban, C., Peimbert, 
M., Torres-Peimbert, S., \& Escalante, V.\ 1998, \mnras, 295, 401 

\bibitem[Ferland et al.(1998)]{Ferland et al. 1998} Ferland, G.~J., Korista, 
K.~T., Verner, D.~A., Ferguson, J.~W., Kingdon, J.~B., \& Verner, E.~M.\ 1998, 
\pasp, 110, 761 

\bibitem[Fich \& Silkey(1991)]{Fich & Silkey 1991} Fich, M., \& Silkey, 
M.\ 1991, \apj, 366, 107 

\bibitem[Fitzgerald(1970)]{Fitzgerald 1970} Fitzgerald, M.~P.\ 1970, \aap, 4, 
234 

\bibitem[Galad{\'{\i}}-Enr{\'{\i}}quez et al.(2000)]{Galadi-Enriquez et al. 
2000} Galad{\'{\i}}-Enr{\'{\i}}quez, D., Trullols, E., \& Jordi, C.\ 2000, 
\aaps, 146, 169 

\bibitem[Gum(1955)]{Gum 1955} Gum, C.~S.\ 1955, \memras, 67, 155 

\bibitem[Herbst(1975)]{Herbst 1975} Herbst, W.\ 1975, \aj, 80, 212 

\bibitem[Israel(1978)]{Israel 1978} Israel, F.~P.\ 1978, \aap, 70, 769 

\bibitem[Kaler(1976)]{Kaler 1976} Kaler, J.~B.\ 1976, \apjs, 31, 

\bibitem[Kimeswenger(2001)]{Kimeswenger 2001} Kimeswenger, S.\ 2001, 
Rev. Mex. Astron. Astrof., 37, 115 

\bibitem[Kingsburgh \& Barlow(1994)]{Kingsburgh & Barlow 1994} Kingsburgh, 
R.~L., \& Barlow, M.~J.\ 1994, \mnras, 271, 257 
 
\bibitem[Krabbe \& Copetti(2005)]{Krabbe & Copetti 2005} Krabbe, A.~C., \& 
Copetti, M.~V.~F.\ 2005, \aap, 443, 981 

\bibitem[Krabbe \& Copetti(2006)]{Krabbe & Copetti 2006} Krabbe, A.~C., \& 
Copetti, M.~V.~F.\ 2006, \aap, 450, 159 

\bibitem[Landolt(1992)]{Landolt 1992} Landolt, A.~U.\ 1992, \aj, 104, 340 

\bibitem[Lauberts et al.(1981)]{Lauberts et al. 1981} Lauberts, A., 
Holmberg, E.~B., Schuster, H.-E., \& West, R.~M.\ 1981, \aaps, 43, 307 

\bibitem[Lindoff(1968)]{Lindoff 1968} Lindoff, U.\ 1968, Arkiv for Astronomi, 5, 
63 

\bibitem[Liu et al.(2000)]{Liu et al. 2000} Liu, X.-W., Storey, P.~J., 
Barlow, M.~J., Danziger, I.~J., Cohen, M., \& Bryce, M.\ 2000, \mnras, 312, 
585 

\bibitem[Martins et al.(2005)]{Martins et al. 2005} Martins, F., Schaerer, D., 
\& Hillier, D.~J.\ 2005, \aap, 436, 1049 

\bibitem[Nordstrom(1975)]{Nordstrom 1975} Nordstrom, B.\ 1975, \aaps, 21, 193 

\bibitem[Peimbert \& Torres-Peimbert(1971)]{Peimbert & Torres-Peimbert 1971} 
Peimbert, M., \& Torres-Peimbert, S.\ 1971, \apj, 168, 413 

\bibitem[Perinotto et al.(2004)]{Perinotto et al. 2004} Perinotto, M., 
Morbidelli, L., \& Scatarzi, A.\ 2004, \mnras, 349, 793 

\bibitem[Rousseau \& Perie(1996)]{Rousseau & Perie 1996} Rousseau, J.~M., \& 
Perie, J.~P.\ 1996, \aaps, 115, 517 

\bibitem[Schmidt-Kaler(1982)]{Schmidt-Kaler 1982} Schmidt-Kaler, T.\ 1982, in 
Landolt-B\"ornstein, New Series, Group VI, Vol.2, ed. K. Schaifers \& H.~H. 
Voigt (Berlin: Springer-Verlag)

\bibitem[Schaerer \& de Koter(1997)]{Schaerer & de Koter 1997} Schaerer, D., \& 
de Koter, A.\ 1997, \aap, 322, 598 

\bibitem[Storey \& Hummer(1995)]{Storey & Hummer 1995} Storey, P.~J., \& 
Hummer, D.~G.\ 1995, \mnras, 272, 41 

\bibitem[Tenorio-Tagle(1979)]{Tenorio-Tagle 1979} Tenorio-Tagle, G.\ 1979, 
\aap, 71, 59 

\bibitem[Vacca et al.(1996)]{Vacca et al. 1996} Vacca, W.~D., Garmany, 
C.~D., \& Shull, J.~M.\ 1996, \apj, 460, 914 

\bibitem[van den Bergh \& Herbst(1975)]{van den Bergh & Herbst 1975} van den 
Bergh, S., \& Herbst, W.\ 1975, \aj, 80, 208 

\bibitem[Vilchez \& Esteban(1996)]{Vilchez & Esteban 1996} Vilchez, J.~M., \& 
Esteban, C.\ 1996, \mnras, 280, 720 

\bibitem[Walborn \& Fitzpatrick(1990)]{Walborn & Fitzpatrick 1990} Walborn, 
N.~R., \& Fitzpatrick, E.~L.\ 1990, \pasp, 102, 379 

\end{thebibliography}
\end{document}